\documentclass[
    ,final          
  ]
  {aipproc}

\layoutstyle{6x9}

\newcommand{\be}{\begin{eqnarray}}
\newcommand{\ee}{\end{eqnarray}}

\begin{document}
\title{Microscopic Black Holes as a Source of Ultrahigh Energy
$\gamma$-rays}
\author{Roberto Casadio}
{address={Dipartimento di Fisica, Universit\`a di
Bologna and I.N.F.N., Sezione di Bologna,
via Irnerio 46, 40126 Bologna, Italy},
email={casadio@bo.infn.it}}
\author{Benjamin Harms}
{address={Department of Physics and Astronomy,
The University of Alabama,
Box 870324, Tuscaloosa, AL 35487-0324, USA},
email={bharms@bama.ua.edu}}
\author{Octavian Micu}
{address={Department of Physics and Astronomy,
The University of Alabama,
Box 870324, Tuscaloosa, AL 35487-0324, USA},
email={micu001@bama.ua.edu}}
\begin{abstract}
We investigate the idea that ultrahigh energy $\gamma$-rays
($E>10\,$TeV) can be produced when charged particles are
accelerated by microscopic black holes.
We begin by showing that microscopic black holes may exist as
remnants of primordial black holes or as products of the
collisions in the large extra dimensions scenario of high energy
cosmic rays with atmospheric particles.
We then solve Maxwell's equations on curved spacetime backgrounds
in 4, 5 and 6 dimensions and use the solutions to calculate the
energy distributions.
From the latter we obtain the black hole parameters needed to
produce the energies of the observed $\gamma$-rays.
\end{abstract}
\maketitle
\section{Introduction}
The existence of cosmological black holes with masses of $10 - 10^8$
solar masses is now an accepted fact.
The existence of microscopic black holes with masses of $1 - 10$ Planck
masses has not been established, but their existence is of great theoretical
interest.
If they do exist, there is the possibility that they could be probed
experimentally to obtain information about quantum gravity.
One possible source of such microscopic black holes is primordial black
holes, which have reached Planck-size masses during the present epoch
through emission of Hawking radiation.
Another possible source is microscopic black hole production.
The recent proposal of the existence of large extra dimensions
\cite{arkani} and the consequent lowering of the fundamental energy scale
to 1~TeV implies that microscopic black holes can be created in
accelerators \cite{dimopoulos} whose center of mass energies are above
the fundamental energy scale.
Microscopic black holes would also be produced in the collisions of
ultrahigh energy cosmic rays with the Earth's atmosphere
\cite{giddings}.
In this talk we describe one of the signatures of black holes impinging
upon the Earth's atmosphere: ultrahigh energy $\gamma$-rays.
As of yet there is no firm evidence that such emissions are occurring
in our atmosphere, but there is some evidence for $\gamma$-rays with
energies greater than 100~TeV at the 1.6~$\sigma$-level from unknown
sources within the galatic plane\cite{bori}.
\par
A charged particle being accelerated by a black hole can produce
$\gamma$-rays with energies in the multi-TeV range before the
particle passes beyond the horizon radius provided that the
curvature gradient of the space around the black hole is large
enough.
Such curvature gradients occur in quantum black holes, black holes
whose masses are of the order the Planck mass.
A calculation taking into account special relativity (but not general
relativity) shows us that to produce $\gamma$-ray energies in the
10~TeV range a single electronic charge would have to be accelerated
by a black hole with a mass equal to five times that of the Planck
mass.
\par
The microscopic black holes needed to produce ultrahigh
$\gamma$-rays may be the remnants of primordial black holes.
Such black holes can be produced by
\begin{itemize}
\item{Inflationary horizon-scale fluctuations}
\item{Density fluctuations at phase transitions and bubble formation
and collapse}
\item{Baryon isocurvature fluctuations on small scales}.
\end{itemize}
Large-mass primordial black holes ($M>10^{15}\,$gm) decaying via Hawking
radiation \cite{hawking} as described by the canonical ensemble in 4
space-time dimensions, $(dM/dt)\sim-M^{-2}$, would have
decayed to a Planck-size mass in the present epoch.
Microscopic black holes produced in this manner could be stable if
quantum gravity effects terminate the decay process.
\par
Copious microscopic black hole production can also occur if large extra
dimensions exist.
In this scenario black hole production can occur as the result of the
collision of particles with total center of mass energy above the
effective Planck scale, which can be as low as the electroweak scale
$m_{ew}\sim 1\,$TeV.
Black holes could thus be produced in collisions of high energy cosmic
rays with the Earth's atmosphere.
As we show in the next Section, such black holes may live long enough
to create ultrahigh $\gamma$-rays even without taking quantum gravity
effects into account.
\section{Black Holes and Large Extra Dimensions}
In a 4-dimensional space-time, a black hole might emerge from
the collision of two particles only if its center of mass energy
exceeds the Planck mass $m_p$ ($l_p$ will denote the Planck length).
In fact, the Compton wavelength $l_M=l_p\,(m_p/M)$ of
a point-like particle of mass $M<m_p$ would be smaller than the
gravitational radius $R_H=2\,G_N\,M=2\,(l_p/m_p)\,M$ and the very
(classical) concept of a black hole would lose its meaning.
However, since the fundamental mass scale is shifted down to
$m_{ew}$ in the models under consideration, black holes with
$M\ll m_p$ can now exist as classical objects provided
\be
l_p\,(m_p/M)\ll R_H\ll L
\ ,
\label{class}
\ee
where $L$ is the scale at which corrections to the Newtonian
potential become effective.
The left hand inequality ensures that the black hole behaves
semiclassically, and one does not need a full-fledged theory of
quantum gravity, while the right hand inequality guarantees that the
black hole is small enough that its gravitational field can depart
from the Newtonian behavior without contradicting present experiments.
\par
The luminosity of a black hole in $D$ space-time dimensions is
given by
\be
{\mathcal L}_{(D)}(M)=
{\cal A}_{(D)}\,\int_0^{\infty}\sum_{s=1}^S\,
n_{(D)}\,(\omega)\,\Gamma_{(D)}^{(s)}(\omega)\,\omega^{D-1}\,
d\omega
\label{dMdt}
\ee
where ${\mathcal A}_{(D)}$ is the horizon
area in $D$ space-time dimensions, $\Gamma_{(D)}^{(s)}$ the
corresponding grey-body factor and $S$ the number of species of
particles that can be emitted.
For the sake of simplicity, we shall approximate
$\sum_s\,\Gamma^{(s)}_{(D)}$ as a constant.
The distribution $n_{(D)}$ is the microcanonical number density
\cite{r1,mfd,thermo}
\be
n_{(D)}(\omega)=C\sum_{l=1}^{[[M/\omega]]}
\exp\left[S_{(D)}^E(M-l\,\omega)-S_{(D)}^E(M)\right]
\label{n}
\ee
where $[[X]]$ denotes the integer part of $X$ and $C=C(\omega)$
encodes deviations from the area law \cite{r1} (in the following
we shall also assume $C$ is a constant in the range of interesting
values of $M$).
\subsection{ADD scenario}
If the space-time is higher dimensional and the extra dimensions
are compact and of size $L$, the relation between the mass of a
spherically symmetric black hole and its horizon radius is changed
to \cite{myers}
\be
R_H\simeq l_{(4+d)}\,
\left({2\,M\over m_{(4+d)}}\right)^{1\over 1+d}
\ ,
\label{R_H<}
\ee
where
$G_{(4+d)}\simeq L^d\,G_N$
is the fundamental gravitational constant in $4+d$ dimensions.
Eq.~(\ref{R_H<}) holds true for black holes of size $R_H\ll L$,
or, equivalently, of mass $M\ll M_c\equiv m_p\,(L/l_p)$.
Since $L$ is related to $d$ and the fundamental mass scale
$m_{(4+d)}$ by \cite{arkani}
\be
L\sim
\gamma^{1+{2\over d}}\, 10^{{31\over d}+16}\,l_p
\ ,
\label{tev}
\ee
where $\gamma\equiv m_{ew}/m_{(4+d)}$,
Eq.~(\ref{class}) translates into
\be
10^{-{31+16\,d\over 2+d}}\,\gamma\,m_p
\sim 10^{-16}\,\gamma\,m_p
\ll M\ll M_c
\ ,
\ee
\par\noindent
where we also used the fact that $d=1$ is ruled out by present
measurement of $G_N$ \cite{arkani} and relatively high values
of $d$ (i.e., $d\sim 6$) seem to be favored
(see, e.g., Refs.~\cite{bounds}).
For $\gamma\sim 1$ (i.e., $m_{(4+d)}\sim m_{ew}\sim 1\,$TeV),
the left hand side above is of order $m_{ew}$ as well.
\par
\begin{figure}
\hspace{-4cm}
\raisebox{7cm}{${\cal L}$}
\includegraphics[width=12cm]{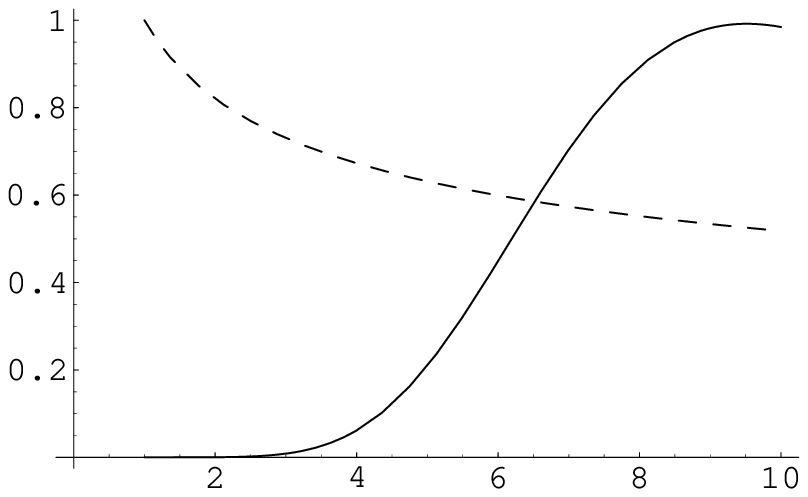}
\hspace{-4cm}${M\over m_{ew}}$
\caption{Microcanonical luminosity (solid line) for a small
black hole with $d=6$ extra dimensions compared to the
corresponding canonical luminosity (dashed line).
Vertical units are chosen such that the canonical luminosity
${\mathcal L}(m_{ew})=1$.}
\label{L_add}
\end{figure}

For $R_H<L$ the Euclidean action
$S_E^<\sim \left({M/m_{ew}}\right)^{(d+2)/(d+1)}$ and
the occupation number density for the Hawking particles in the
microcanonical ensemble is given by
\be
n_{(4+d)}(\omega)\sim\sum_{l=1}^{[[M/\omega]]}\,
e^{\left({M-l\,\omega\over m_{ew}}\right)^{d+2\over d+1}
-\left({M\over m_{ew}}\right)^{d+2\over d+1}}
\ .
\label{n_add}
\ee
In 4 dimensions one knows that microcanonical corrections
to the luminosity become effective only for $M\sim m_p$,
therefore, for black holes with $M\gg m_{ew}$ the luminosity
(\ref{dMdt}) should reduce to the canonical result
\cite{bc1,bc2,emparan}.
However, in $4+d$ dimensions, such corrections are not entirely
negligible for $M\sim m_{ew}$.
The decay rate corresponding to the number density (\ref{n_add})
as calculated from Eq.~(\ref{dMdt}) is exhibited for $d=6$ in
Fig.~\ref{L_add}.
For lower values of $d$ the peak in the microcanonical luminosity
shifts to smaller values of $M$, thus approaching the 4-dimensional
canonical decay rate (also shown for comparison in Fig.~\ref{L_add}).
In all cases, the microcanonical luminosity becomes smaller
for $M\sim m_{ew}$ than it would be according to the
canonical luminosity, which makes the life-time of the
black hole somewhat longer than in the canonical
picture.
In particular, for $d=6$ one finds
\be
\left.{dM\over dt}\right|_{M\sim m_{ew}}
\simeq -10^{-10}\,{\mathcal L}_{(10)}^H
\sim -10^{17}\,{{\rm TeV}\over {\rm s}}
\ .
\ee
A black hole would therefore evaporate very quickly
\cite{dimopoulos} down to $\sim 6\,m_{ew}$.
Then, its life-time is dominated by the time it would
take to emit the remaining $\Delta M\sim 5\,$TeV,
before it reaches $1\,m_{ew}$, which is approximately
\be
T\sim \left({dM\over dt}\right)^{-1}\,\Delta M
\sim 10^{-17}\,{\rm s}
\ .
\label{T}
\ee
The above relatively long time also takes into account
the dependence of the grey-body factor $\Gamma_{(4+d)}^{(s)}$
on $d$ (for the details see Ref.~\cite{lhc}).
Without a full-fledged theory of quantum gravity, nothing
can be safely stated as $M$ becomes less than $m_{ew}$.
However, by extrapolating the microcanonical behavior,
one would conclude that the evolution then proceeds roughly
according to the usual exponential law of radiative
decay \cite{mfd}.
\subsection{RS scenario}
In order to study this case, we shall make use of the
solution given in Ref.~\cite{maartens}.
This is one of the few known metrics on the brane which might
represent such a case in the context of the RS scenario (for
more candidates see Ref.~\cite{cfm}).
Such a solution has the Reissner-Nordstr\"om form
\be
-g_{tt}={1\over g_{rr}}=
1-2\,{M\,l_p\over m_p\,r} +Q^2\,{l_p^2\over r^2}
-q\,{m_p^2\,l_p^2\over m_{(5)}^2\,r^2}
\ ,
\ee
and the (outer) horizon radius is given by
\be
R_H=l_p\,{M\over m_p}\,
\left[1+\sqrt{1-Q^2\,{m_p^2\over M^2}
+{q\,m_p^4\over M^2\,m_{(5)}^2}}\right]
\ ,
\label{RH_RS}
\ee
where $m_{(5)}\sim m_{ew}$ is the fundamental mass scale and
$q$ represents a (dimensionless) tidal charge.
The latter can be estimated on dimensional grounds as
\cite{bc2,maartens}
$q\sim\left({m_p\over m_{ew}}\right)^\alpha\,{M\over m_{ew}}$
and for $\alpha>-4$ the tidal term $\sim 1/r^2$ dominates over the
4-dimensional potential $\sim 1/r$ (as one would expect for
tiny black holes).
From Eq.~(\ref{RH_RS}) with $Q=0$ and $\alpha>-4$ one
obtains
\be
R_H\simeq
l_p\,\left({m_p\over m_{(5)}}\right)^{1+{\alpha\over 2}}\,
\sqrt{M\over m_{(5)}}
\ ,
\ee
since the tidal term $q$ dominates for both $M$ and
$m_{(5)}\ll m_p$, and one must still have Eq.~(\ref{class}).
With one warped extra dimension \cite{RS}, the
length $L$ is just bounded by requiring that Newton's law not be
violated in the tested regions, since corrections to the $1/r$
behavior are of order $(L/r)^2$.
This roughly constrains $l_p<L<10^{-3}\,$cm.
Hence the allowed masses are, according to
Eq.~(\ref{class}),
\be
\left({m_{(5)}\over m_{p}}\right)^{{\alpha\over 3}}
\ll {M\over m_{(5)}}
\ll\left({L\over l_p}\right)^{2}\,
\left({m_{(5)}\over m_p}\right)^{2+\alpha}
\ .
\ee
In particular one notices that black holes with
$M\sim m_{(5)}\sim m_{ew}$ could exist only if
the following two conditions are simultaneously satisfied
\be
\alpha\ge 0
\ \ \ \ {\rm and}
\ \ \ \ {L\over l_p}\gg
\left({m_p\over m_{ew}}\right)^{3+\alpha\over 3}
\ .
\label{exist}
\ee
\par
The luminosity is now given by,
for the limiting case $\alpha=0$ and taking into account the second
condition in Eq.~(\ref{exist}),
\be
{\mathcal L}_{(4)}<
10^{-9}\,{M\over m_{ew}}\,{{\rm TeV}\over{\rm s}}
\ ,
\ee
which yields an exponential decay with typical life-time
$T>10^9\,$s.
\subsection{Production of Black Holes by Cosmic Rays}
Since the Planck mass in $4+d$ dimensions can be as small as 1~TeV,
black holes can be created in cosmic ray interactions with particles
in the Earth's atmosphere (or any other source of matter within the
galaxy) in the processes
\be
\begin{array}{l}
p + p \to {\rm BH} + X^{++}
\\
\\
\nu + N \to {\rm BH} + X
\ .
\end{array}
\nonumber
\ee
The cross section for the production of a black hole with mass
$M$ in such processes is given to a good approximation by
$\sigma \simeq \pi\, R^2_{H}$, where
\be
R_H={1\over{\sqrt{\pi}\,m_{(4+d)}}}\,\left[
{M\over{m_{(4+d)}}}\,
\left({8\,\Gamma({{d+3}\over{2}})
\over{d+2}}\right)\right]^{{1\over{d+1}}}
\ ,
\ee
and we recall that $m_{(4+d)}\sim m_{ew}$.
For proton and neutrino energies above $\sim 10^{8}\,$TeV black hole
production will dominate over the production of standard model
particles~\cite{ring}.
About 100 black holes per year are created for the whole surface of
the Earth by the $p+p$ process~\cite{giddings}, while the $\nu + N$
process creates one black hole per year~\cite{ring}.
\section{Acceleration of a Charged Particle by a Neutral Black Hole}
\subsection{4-Dimensional Case}
\begin{figure}
\hspace{-3cm}
\raisebox{7cm}{${1\over q^2}\,{dE\over d\omega}$}
\hspace{-1cm}
\includegraphics[width=14cm]{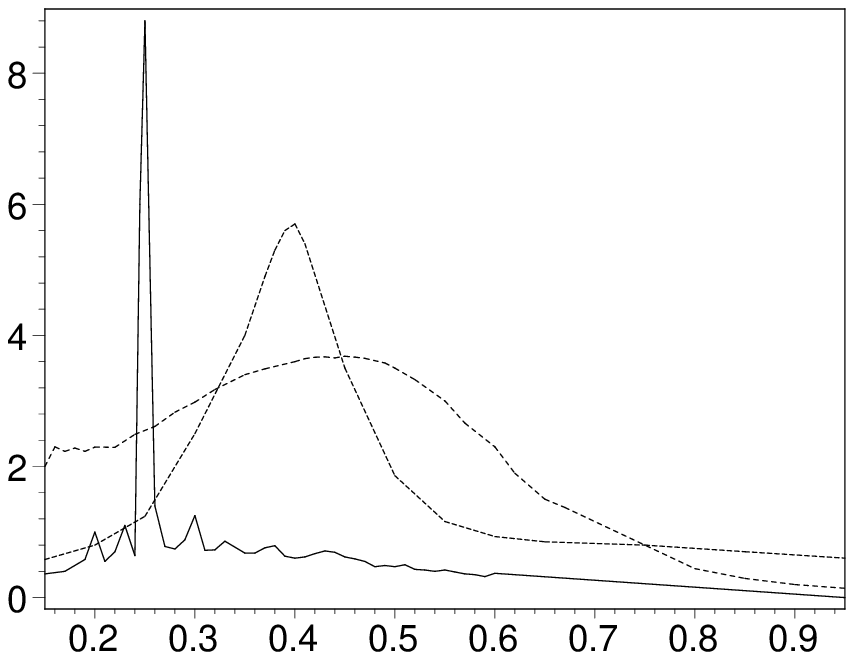}
\hspace{-10cm}
\hspace{6cm}${\omega\,R_H}$
\caption{Energy distribution for a particle of unit charge
falling into a 4-D black hole of mass $M$ (solid line).
Same distribution (scaled by 100) for a 5-D black hole
(dashed line) and (scaled by 1000) for a 6-D black hole
(dotted line).
Note that secondary peaks are due to limited numerical
precision.}
\label{Edist}
\end{figure}

To describe the radiation a charged particle being accelerating by
a neutral microscopic black hole will produce, Maxwell's equations
must be solved on a curved spacetime background.
The perturbed electromagnetic tensor elements $f_{\mu\nu}$ are
determined from the relation
\be
\left(\sqrt{-g}\,f^{\mu\nu}\right)_{,\nu}
=4\,\pi\,\sqrt{-g}\,j^\mu
\ ,
\label{max}
\ee
where $j^\mu$ is the current associated with the falling charge.
In 4 dimensions the Schwarzschild metric is spherically symmetric
(in this Section the masses are in units of length, i.e.
$G_N=1$),
\be
ds^2=-\left(1+{2\,M\over r}\right)\,dt^2
+\left(1+{2\,M\over r}\right)^{-1}\,dr^2
+ r^2\,\left(d\theta^2+\sin^2\theta\,d\phi^2\right)
\ ,
\ee
and, in order to solve the set of equations in Eq.~(\ref{max}) the
perturbations $f_{\mu\nu}$ are conveniently expanded in tensor
harmonics \cite{zerilli}.
Using the field equations all of the tensor elements $f_{\mu\nu}$ can be put
in terms of a single element, say $f_{12}$.
After taking the Fourier transform
(${\partial\over{\partial t}}\to i\omega)$ the remaining element
satisfies the equation
\be
\displaystyle{\frac{d^2\,f_{lm}}{dr_*^{2}}}
+\left\{\omega^2-{\rm e}^\nu\left[\displaystyle{
\frac{l(l+1)}{r^2}}\right]\right\}f_{lm}=
2\;\sqrt{l+1/2}\;{\rm e}^{i\omega T}\;{\rm e}^\nu\left[{\rm e}^\nu\;u
-\displaystyle{\frac {d}{dr}({\rm e}^\nu w)}\right]
\label{flm}
\ee
where $r_*$ is the standard ``turtle'' coordinate,
\be
f_{lm}={\rm e}^{\nu}\,f_{12}
\ ,
\ee
\be
{\rm e}^\nu&=&1-{\displaystyle\frac{2\,M}{r}}\ ,\ \ \ \
u={\displaystyle\frac{q\,{\rm e}^{-\nu}}{r^2}}\ ,\ \ \ \
w={\displaystyle\frac{q}{l(l+1)}}{\displaystyle \frac
{dT}{dr}}
\ ,
\\
T&=&-4\,M\,\sqrt{r/(2\,M)}
-(4/3)\,M\,\left[r/(2\,M)\right]^{3/2}
-2\,M\,\ln\left(\sqrt{r/(2\;M)}-1\right)
\nonumber
\\
&&
+2\,M\,\ln\left(\sqrt{r/(2\;M)}+1\right)
\ .
\ee
The charge on the infalling particle is $q$ and $T$ is the time
for the particle to fall from $\infty$ to the point $r$.
The equation for $f_{lm}$ cannot be solved analytically, but
values for the amplitude can be obtained numerically using the
Green's function $G(r,r')$.
The solution of Eq.~(\ref{flm}) is given by
\be
f_{lm}=\int G(r,r')\,S(r')\,dr'
\ee
where $S(r)$ is the source term on the right hand side of
Eq.~(\ref{flm}).
The energy distribution averaged over the
complete solid angle is
\be
\left\langle{\displaystyle{\frac{dE}{d\omega}}}\right\rangle
=\displaystyle{\frac{l(l+1)}{2\;\pi}}f_{l0}\;f^*_{l0}
\ .
\ee
This distribution is plotted as a solid line in Fig.~\ref{Edist}
for unit charge $q$ and black hole mass $M$ ($R_H=2\,M$).
The energy of the radiation is
\be
\Delta E
\simeq
\displaystyle{\frac {C_{(4)} q^2}{M}}
\ ,
\label{En}
\ee
where $C_{(4)}$ is the area under the curve.
To produce a 100~TeV photon from a singly charged particle ($q = e$)
the black hole would have to have a mass of approximately $10\,m_p$.
\subsection{4+d-Dimensional Case}
For the case of large extra dimensions the invariant line
element is taken to be of the form
\be
ds^2&=&
-\left[1-\displaystyle\left({R_H^2\over r^2+\sum y_i^2}\right)^{d+1\over 2}
\right]\,dt^2
+\left[1-\displaystyle\left({R_H^2\over r^2+\sum y_i^2}\right)^{d+1\over 2}
\right]^{-1}\,dr^2
\nonumber
\\
&&+r^2\,\left(d\theta^2+\sin^2\theta\,d\phi^2\right)
+\sum dy_i^2
\ ,
\ee
where the $y_i$'s are the coordinates of the $d$ extra dimensions.
We assume in this case that the electromagnetic radiation is confined
to a 4-dimensional brane embedded in the $4+d$-dimensional space.
Thus the $y_i$'s are set to zero in $g_{00}$ and $g_{11}$, and the
harmonic tensor has the same form as in Ref.~\cite{zerilli}, but with
extra rows and columns of zeros for the extra dimensions.
The energy distributions for the $d=1,2$ cases are given in
Fig.~\ref{Edist} together with the 4-dimensional case.
The corresponding energy expressions are of the same form as for
the 4-dimensional case [Eq.~(\ref{En})].
It is clear from the graphs that heights of the peaks decrease, that
the peaks broaden and that the centers of the peaks shift to larger
values of the frequency $\omega$ with increasing number of extra
dimensions.
The expressions used for the metric tensor elements for the
$d > 0$ cases are approximations which are valid if $R_H \gg L$. The energy radiated in these cases is given by
\be
\Delta E\simeq
\displaystyle{\frac {C_{(D)}\, q^2}{R_H}} \ \ \ \ \ \ \
D= 5,\, 6
\ ,
\label{En2}
\ee
where the $C_{(D)}$'s are twice the areas under the corresponding
curves.
\par
The approximation used to obtain the expressions for the 5- and
6-dimensional cases is not satisfactory for $R_H \ll L$.
The latter is the more realistic case for microscopic black holes,
but there is no known metric which suitably describes this case.
Therefore the problem of obtaining black hole parameters from the
observed energies of the $\gamma$-rays remains to be solved.
\begin{theacknowledgments}
This work was supported in part by the U.S. Department of Energy
under Grant no.~DE-FG02-96ER40967.

\end{theacknowledgments}
%
%

%
%
\end{document}